\documentclass[preprint,aps,tightenlines,floatfix,showpacs]{revtex4}
\usepackage{graphics}
\usepackage{bm}
\usepackage{amsmath}
\usepackage{amssymb}
\begin{document}
\preprint{}
\title{Is friction responsible for the reduction of fusion rates far below
  the Coulomb barrier?}
\author{B. G. Giraud}
\email{giraud@spht.saclay.cea.fr}
\affiliation{Service de Physique Th\'eorique, DSM, CEA Saclay, F-91191
  Gif-sur-Yvette, France}
\author{S. Karataglidis}
\email{kara@physics.unimelb.edu.au}
\affiliation{School of Physics, University of Melbourne, 
  Victoria 3010, Australia}
\author{K. Amos}
\email{amos@physics.unimelb.edu.au}
\affiliation{School of Physics, University of Melbourne, 
  Victoria 3010, Australia}
\author{B. A. Robson}
\email{Brian.Robson@anu.edu.au}
\affiliation{Department of Theoretical Physics, Research School of
  Physical Sciences and Engineering, The Australian National University,
  Canberra, Australian Capital Territory 0200, Australia}
\date{\today}
\begin{abstract}
  The  fusion of  two interacting  heavy ions  traditionally  has been
  interpreted in terms  of the penetration of the  projectile into the
  target.  Observed   rates  well   below  the  Coulomb   barrier  are
  considerably lower than estimates obtained from penetration factors.
  One  approach  in the  analysis  of  the  data invokes  coupling  to
  non-elastic  channels  in  the  scattering  as  the  source  of  the
  depletion. Another is to analyze those data in terms of tunneling in
  semi-classical models,  with the  observed depletion being  taken as
  evidence  of  a  ``friction''  under the  barrier.  A  complementary
  approach is to  consider such tunneling in terms  of a fully quantal
  model.  We  investigate  tunneling  with  both  one-dimensional  and
  three-dimensional models in a  fully quantal approach to investigate
  possible  sources for  such a  friction. We  find that  the observed
  phenomenon may not  be explained by friction. However,  we find that
  under certain conditions tunneling  may be enhanced or diminished by
  up  to  50\%, which  finds  analogy  with  observation, without  the
  invocation of a friction under the barrier.
\end{abstract}
\maketitle

\section{Introduction}

Fusion  reactions  near the  Coulomb  barrier  are  a stimulating  and
challenging  subject in  nuclear physics,  especially given  that most
nucleosynthesis  processes  in  the   Big  Bang  and  in  the  stellar
environment  fall  into this  category.  Theoretical understanding  is
necessary  given that  most of  these reactions,  which are  needed as
input in nucleosynthesis models, may not be measured in the laboratory
due to the very small cross sections involved.

For  reactions  far  below  the  Coulomb barrier  (see,  for  example,
\cite{Ha03,Ji04}),  measured fusion  cross  sections are  considerably
lower compared to their  estimates from penetration factors. Normally,
the conjecture is  that an energy loss has  occurred under the Coulomb
barrier. Intuitively,  that loss may  be understood as  a ``friction''
\cite{Ca81}, accounting  for coupling  to other reaction  channels. An
alternative  postulate  (see \cite{Ha03},  for  example)  is that  the
nucleus-nucleus  optical potential  involved in  fusion  processes may
require a  much larger diffuseness  than that for  elastic scattering.
But that  approach may be problematic  given that the  coupling of the
nonelastic  and elastic  channels in  the  nucleus-nucleus interaction
should  be specified  self-consistently. The  role of  breakup  in the
depletion of  fusion has  also been investigated  \cite{Da02}, wherein
fusion involving weakly bound nuclei  may be diminished by up to 35\%.
Herein, we  investigate the tunneling  hypothesis and the notion  of a
``friction''.

The invocation of a friction is  a result of the use of semi-classical
models.  It  is  an  alternative  to  the  purely  quantal  nature  of
tunneling. The kinetic  energy under a barrier should  be negative and
time under the barrier must be made imaginary, or at least complex, to
compensate for the classical anomaly.  That is essential if a position
and velocity  are to  be used  as measures of  the propagation  of the
fusing ions  under the  barrier. In such  an analytic  continuation of
classical physics  to complex trajectories  and complex time,  can one
then contemplate a  random Langevin force to describe  friction? For a
simpler understanding  of the (quantal)  real time processes,  a fully
quantal model of tunneling is necessary.

Herein  we shall  not  follow  the approach  of  Caldeira and  Leggett
\cite{Ca81} (see also \cite{Wi82,Ca82})  in which the tunneling degree
of freedom  is coupled  to a bath  of harmonic oscillators.  From that
approach, a conclusion  was drawn that a loss  of transmission occurs.
But recent  studies \cite{Fr95,Fr96,Lu03}  show that for  some chaotic
potentials, barrier  penetration in fact  is enhanced. Thus we  seek a
more  pedagogical  approach recognizing  that,  if Langevin  processes
exist to account for  friction, the effective potential experienced by
the  tunneling particle  will not  be smooth.  Thus we  wish  to study
tunneling through rough potentials \textit{in real time}.

Herein, to facilitate such an investigation, we construct models where
wave packets  are prepared far from  the barrier. These  will be broad
packets having  few (if  any) components with  energy higher  than the
barrier. Such  packets are boosted toward  the barrier and  we use the
time-dependent  Schr\"odinger  equation  (TDSE)  as  the  equation  of
motion.   \textit{A  priori},   we  shall   use   two  (nonequivalent)
approaches, namely
\begin{description}
  \item[(i)] space fluctuations of a time independent barrier, and
  \item[(ii)] time fluctuations of a spatially smooth barrier.
\end{description}
The potentials of  case (i) may induce enough  incoherence in the wave
propagation  to trigger some  localization \cite{An58}  so diminishing
the  transmission.  Either  case  may  represent  couplings  to  other
channels.

The paper  is arranged as  follows. In Section~\ref{base}  we describe
our one-dimensional reference model  (base potential). We consider the
effects  of   space  fluctuations  in   Section~\ref{space}  while  in
Section~\ref{time}  we  return  to  the  base  potential  and  instead
consider   modifications   to  it   by   fluctuations   in  time.   In
Section~\ref{radial}    we   solve    the    radial   time-independent
Schr\"odinger equation  with an ensemble  of square wells to  define a
statistical average of transmission.  By this means the more realistic
situation of a Coulomb barrier surrounding an attractive well in three
dimensions  is  taken into  account.  Discussion  and conclusions  are
presented in Section~\ref{tisdone}.

\section{Reference model}
\label{base}

Our one-dimensional model assumes spatially even barriers of the type
\begin{equation}
  V( x, t ) = v(t) \exp\left( -2 \omega x^2 \right)
  \label{barrier}
\end{equation}
for which the TDSE is
\begin{equation}
  i\hbar \frac{\partial \Psi}{\partial t} = \left\{ - \frac{ \hbar^2
  }{ 2M }\frac{ \partial^2 }{ \partial x^2} + V(x,t) \right\} \Psi .
\end{equation}
Arbitrarily  we have  chosen $\omega  =  0.5$ and,  for our  reference
model, $v(t) = 1$.

\subsection{Initial conditions, scales, parities}

The  problem  could  be  intractable  as there  are  four  conflicting
considerations, namely
\begin{enumerate}
\item Gaussian  wave packets, or quasi-Gaussian ones,  are required to
  maintain   an  analogy   of  classical   particles   with  maximally
  well-defined positions and velocities as far as is possible. But
\item under the barrier the wave will certainly not be Gaussian and at
  best one might observe probability bumps. Then
\item wave packets must be  broad enough to avoid excessive zero-point
  energies, but
\item the same packets, or their bumps if any, should be narrower than
  the width of the barrier if the particles are to be localized within
  the barrier.
\end{enumerate}
These problems will be addressed as necessary below.

While dimensionless quantities have been used in the calculations, the
results  of which will  be presented,  time and  length scales  may be
inferred by considering typical  values for the tunneling systems. Let
$V_{\text{max}}$ denote  a typical maximal  height of the  barrier. We
select a  time unit $\Delta t  = \hbar/ V_{\text{max}}$  such that the
potential  $V(x,t)\Delta  t/  \hbar$  occurring in  the  dimensionless
Schr\"odinger   equation  is   of   order  unity.   A  typical   value
$V_{\text{max}} = 60$~MeV gives  $\Delta t \simeq 10^{-23}$~sec. Light
ion problems,  with concomitantly  lower barriers, can  induce greater
time scales  with values $\sim  10^{-22}$~sec. possible. With  $M$ the
(reduced) mass  of the packet,  we select a  length scale $\Delta  x =
\sqrt{\hbar \Delta t/M}$ such that the coefficient $\hbar \Delta t/[ M
(\Delta x)^2  ]$ for the kinetic  energy operator is also  of order 1.
With mass number  $\simeq 10$ and $\Delta t  \simeq 10^{-22}$~sec., we
obtain $\Delta x \simeq 1$~fm. Then, for convenience, we choose $\hbar
= M = 1$.

Given  the precise  determination of  the energies  of  projectiles in
experiment,  realistic wave  packets must  be  initiated significantly
broader than  the barrier.  Our even barrier  [Eq.~(\ref{barrier})] is
centered  at   the  origin   with  a  width   of  order  1.   For  the
one-dimensional problem, we choose an initial wave packet of the form
\begin{equation}
  \Psi_l(x,0) = \pi^{-1/4} \exp\{-ax^2 - bx - c\}
  \, ,
\end{equation}
where  the initial  parameters are  $a_0 =  1/(2\lambda_0^2)$,  $b_0 =
3/\lambda_0  -  iK$,  and  $c_0  =  (\ln  \lambda_0)/2  +  9/2$,  with
$\lambda_0 = 5$ being the initial width of the packet. The momentum of
the packet is given by $K$ and the subscript indicates that the packet
moves from  the left  (negative values of  x). With these  choices the
initial packet also has the form
\begin{equation}
  \Psi_l(x,0) = \frac{\pi^{-1/4}}{\sqrt{5}} \exp\left[ - \frac{( x +
  15 )^2 }{50} + iKx \right] \, .
  \label{pack}
\end{equation}
Typically, $0.3 \leq  K \leq 1.3$ whence the  kinetic energy ($K^2/2$)
is well below the height of the barrier. With initial momentum $K \sim
1$ and similar orders of magnitude for all parameters, the packet will
collide  with  the  barrier   typically  at  times  $\sim  10/K$,  the
penetration reaching  its peak at  $\sim 15/K$, and  full transmission
and reflection should be complete by $\sim 30/K$.

One  may also  reverse the  sign  of $b$  and so  obtain the  solution
$\Psi_r(x,t)$, the  mirror image of $\Psi_l(x,t)$. The  even parity of
$V(x,t)$ makes it possible to solve the TDSE for both the even and odd
waves, $\Psi_+(x,t)$ and $\Psi_-(x,t)$, where
\begin{equation}
\Psi_{\pm}(x,t) = \frac{1}{\sqrt{2}}\left( \Psi_l \pm \Psi_r \right)\;
,
\end{equation}
In principle,  such a change should  also initiate a change  in $c$ to
retain   the  normalization  of   $\Psi_\pm(x,t)$.  But   the  overlap
$\left\langle \Psi_l | \Psi_r  \right\rangle$ is small, $< e^{-9} \sim
10^{-4}$ for  $K = 0$  and $\ll  10^{-6}$ for $K  > 0.5$, so  that the
effect  can be  neglected. Nevertheless,  as  a check,  we maintain  a
comparison  between  the direct  solution  $\Psi_l(x,t)$  and the  sum
$\left[ \Psi_+(x,t) + \Psi_-(x,t) \right] / \sqrt{2}$.

\subsection{Results}

Taking  the  Hamiltonian as  $H  =  -\partial^2/(  2\partial x^2  )  +
e^{-x^2}$, the modulus
\begin{figure}
  \scalebox{0.6}{\includegraphics*{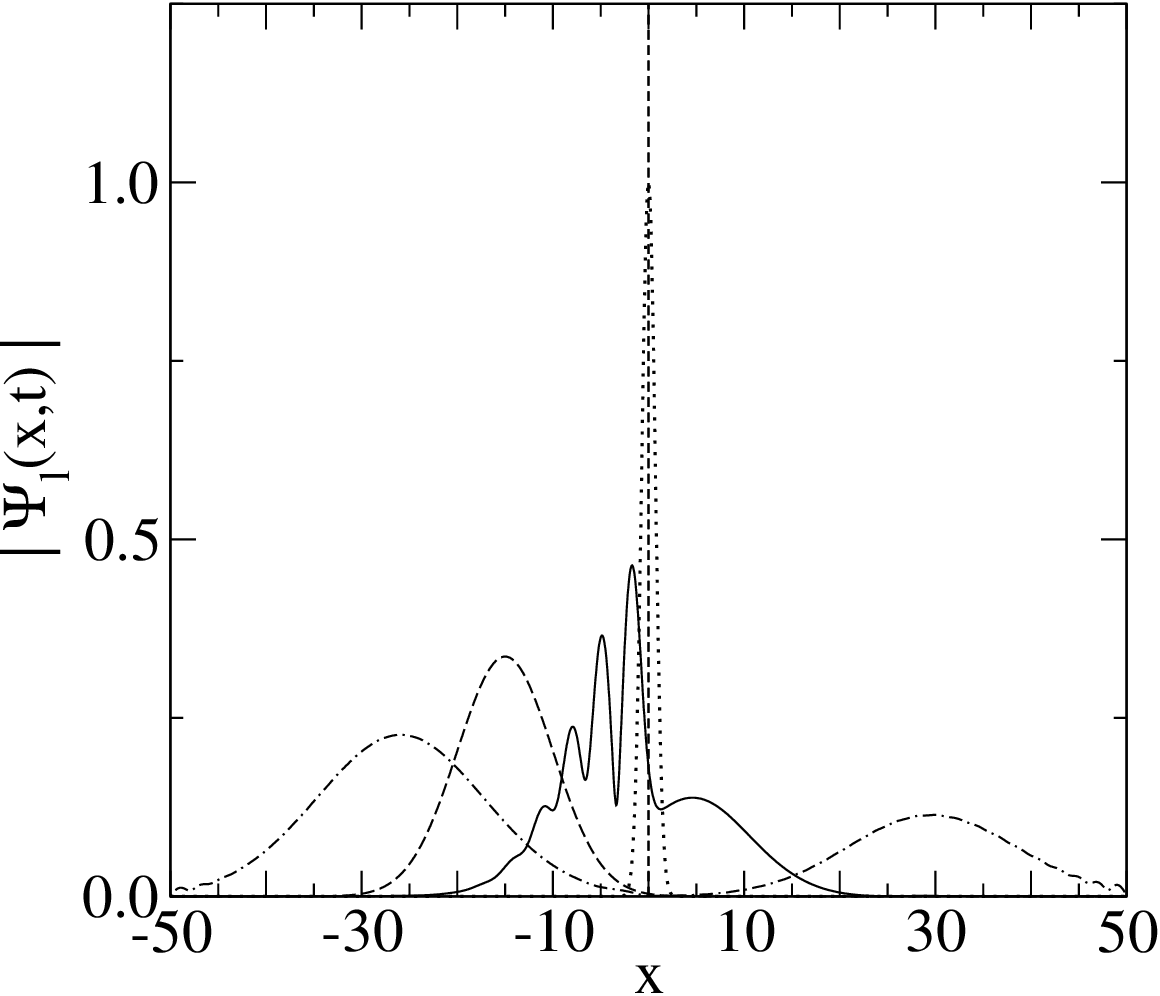}}
  \caption{\label{packets} Modulus  of the wave  packet $| \Psi_l(x,t)
  |$  at  times  $t =  0$  (dashed  line),  18  (solid line),  and  40
  (dot-dashed line). The potential is portrayed by the dotted line.}
\end{figure}
$| \Psi_l(x,t) |$ of the wave packet is shown in Fig.~\ref{packets} at
times $t = 0$, 18, and 40,  for the momentum $K = 1.06$. The potential
is  frozen in  time with  $v(t)  = 1$.  The energy  is 0.571834  which
includes the zero point kinetic energy (0.01) and a small contribution
from the potential energy (0.000034); the latter due to the overlap of
the tails of the potential and  initial wave packet. The energy of the
colliding packet is slightly above half the barrier height.

The difference between  the negative and positive sides  of the packet
at $t  = 18$ is most telling.  The modulus for negative  values of $x$
shows  clearly the  interference  between the  incoming and  reflected
waves while that on the  positive-$x$ side exhibits a good fraction of
the transmitted  packet. At $t =  40$ we clearly  observe two distinct
\textit{Gaussian}   packets  corresponding   to   the  reflected   and
transmitted waves.  The shapes of the reflected  and transmitted waves
have been effectively restored to the original Gaussian shape, with no
memory of the interaction with the potential.

\begin{figure}
  \scalebox{0.6}{\includegraphics*{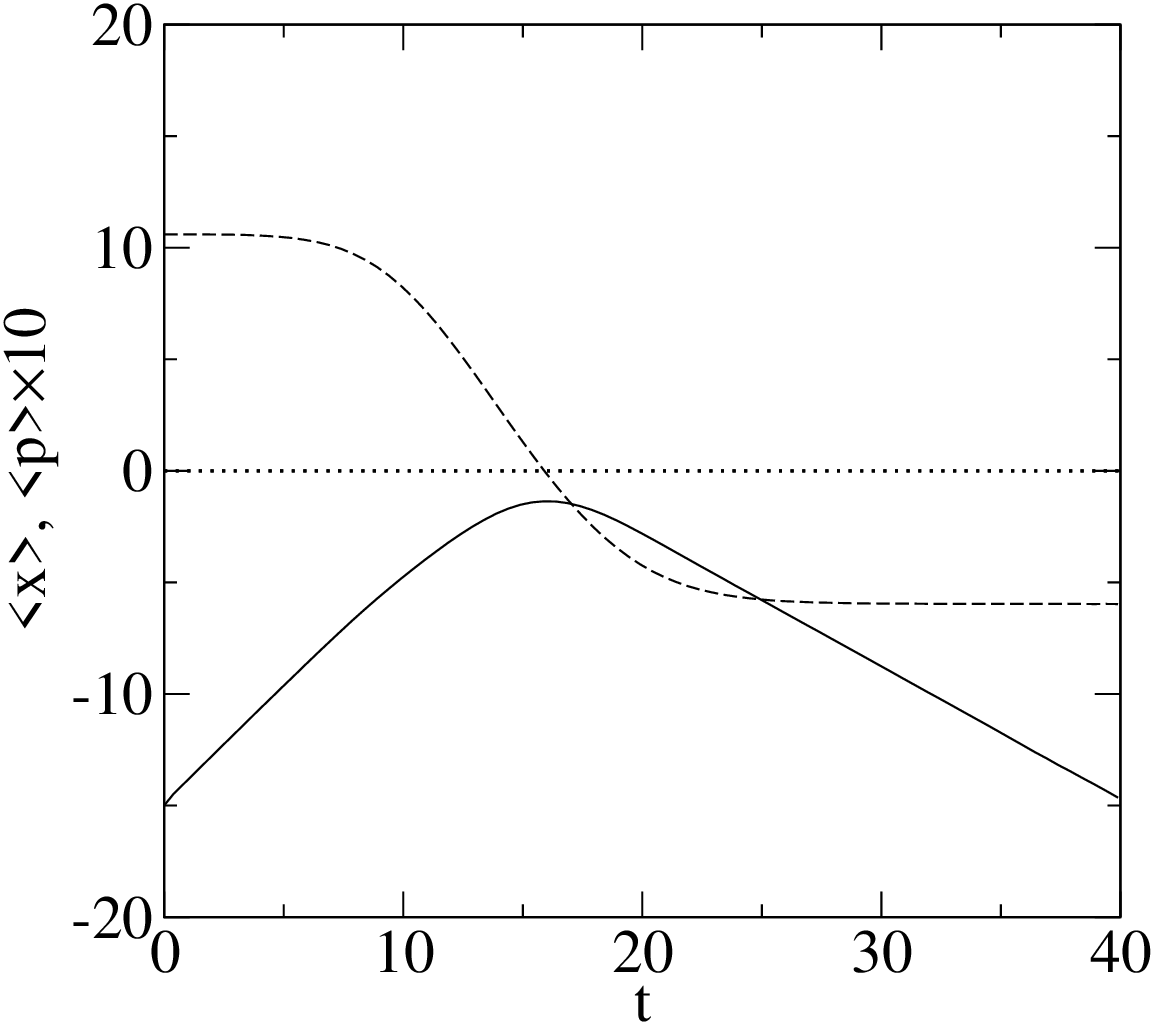}}
  \caption{\label{xp}  The average position  (solid line)  and average
    momentum (dashed line)  of the wave packet as  a function of time.
    The initial momentum is $K = 1.06$. To facilitate the display, the
    momentum has been multiplied by a factor of 10.}
\end{figure}

In Fig.~\ref{xp} the average position
$\left\langle  x \right\rangle$ and  average momentum  $\left\langle p
\right\rangle$ of the wave packet  as a function of time are displayed
as the  solid and dashed  lines respectively. At  $t = 0$,  the packet
center starts from $x = -3\lambda_0 = -15$ with an initial momentum of
$K = 1.06$. The interaction with  the barrier begins at $t \sim 12$ as
denoted by the change in the momentum. The packet slows down and stops
at $t \sim 15$ after which  80\% of it is reflected. The final average
momentum may be estimated to be  $-0.64 = -1.06 \times 0.6$ due to the
transmission of  $\sim 20\%$  of the wave.  This is comparable  to the
actual value of the momentum $\left\langle p \right\rangle = -0.60$ at
$t = 40$.

\begin{figure}
  \scalebox{0.6}{\includegraphics*{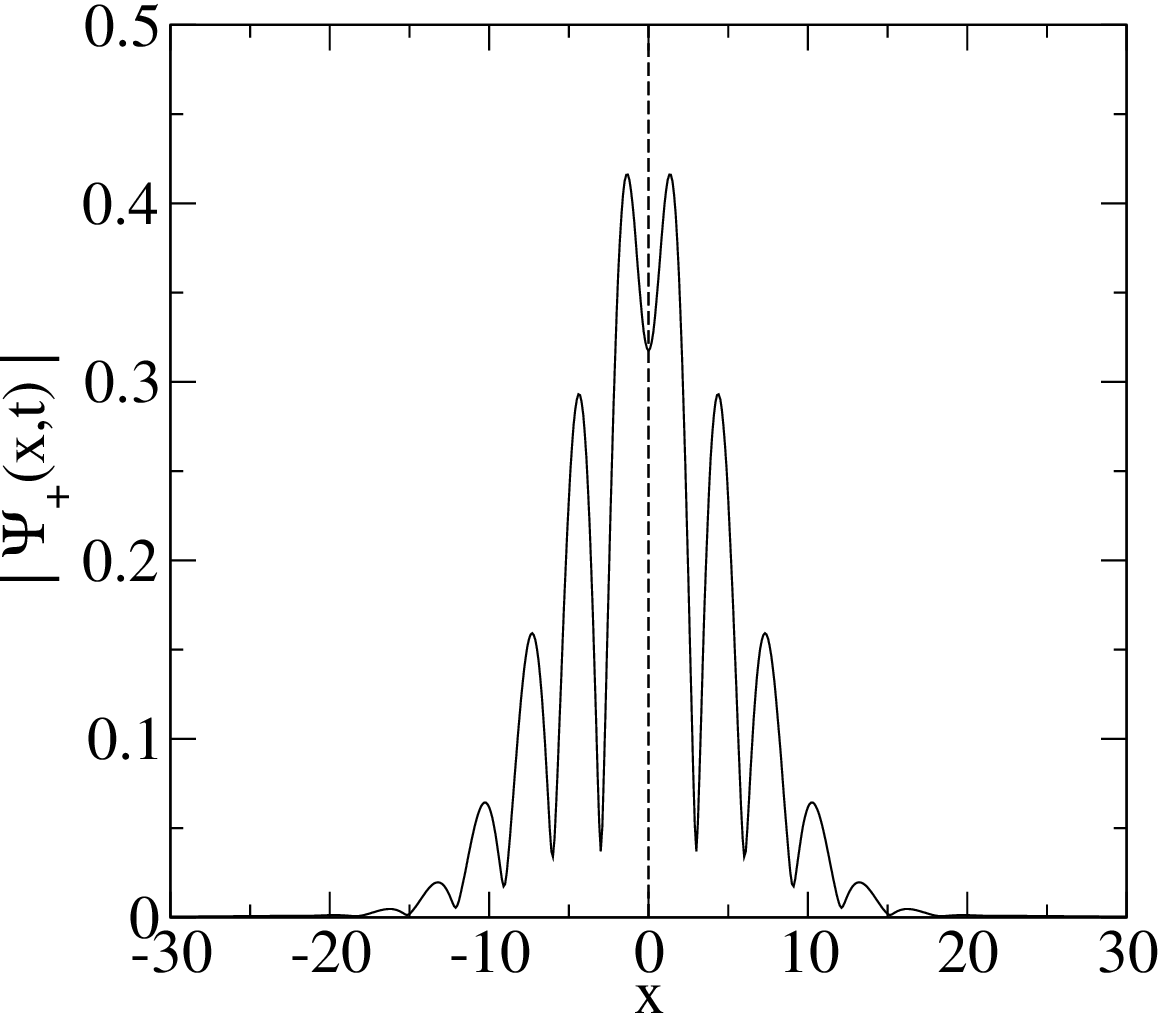}}
  \caption{\label{even}  $\left|  \Psi_+   \right|$  at  $t  =  15.75$
    corresponding to maximal presence under the barrier.}
\end{figure}

In seeking  evidence for  a resonance behavior  under the  barrier, we
also investigated  the behavior of the  even and odd  solutions of the
TDSE. We show in Fig.~\ref{even} the absolute value of the even
solution $\Psi_+$  at $t  = 15.75$ and  for $K =  1.06$. Approximately
these conditions  correspond to  the situation of  maximum interaction
with the  barrier. Note  there is no  peak of $\left|  \Psi_l \right|$
under the barrier,  and $\Psi_-$ has a node at $x  = 0$. Therefore, we
sought to  create a  ``bump under the  barrier'' for  some combination
$\Psi_{\text{mix}}  =   \Psi_+  \cos{\theta}  +   \Psi_-  e^{i\varphi}
\sin{\theta}$. The propagation of this  mixed wave is then analyzed in
classical terms of a bump position and momentum. We could not find any
combination of $\theta$  and $\varphi$ giving such a  bump. This stems
from  the strong  positive  curvature of  $\Psi_+$  at $x  = 0$  while
$\Psi_-$ has no  curvature at its node. Hence,  for any given $\theta$
both $\Psi_{\text{mix}}$ and its  second derivative have the same sign
at $x = 0$. No bump can be found as such would require opposite signs.

\begin{figure}
  \scalebox{0.6}{\includegraphics*{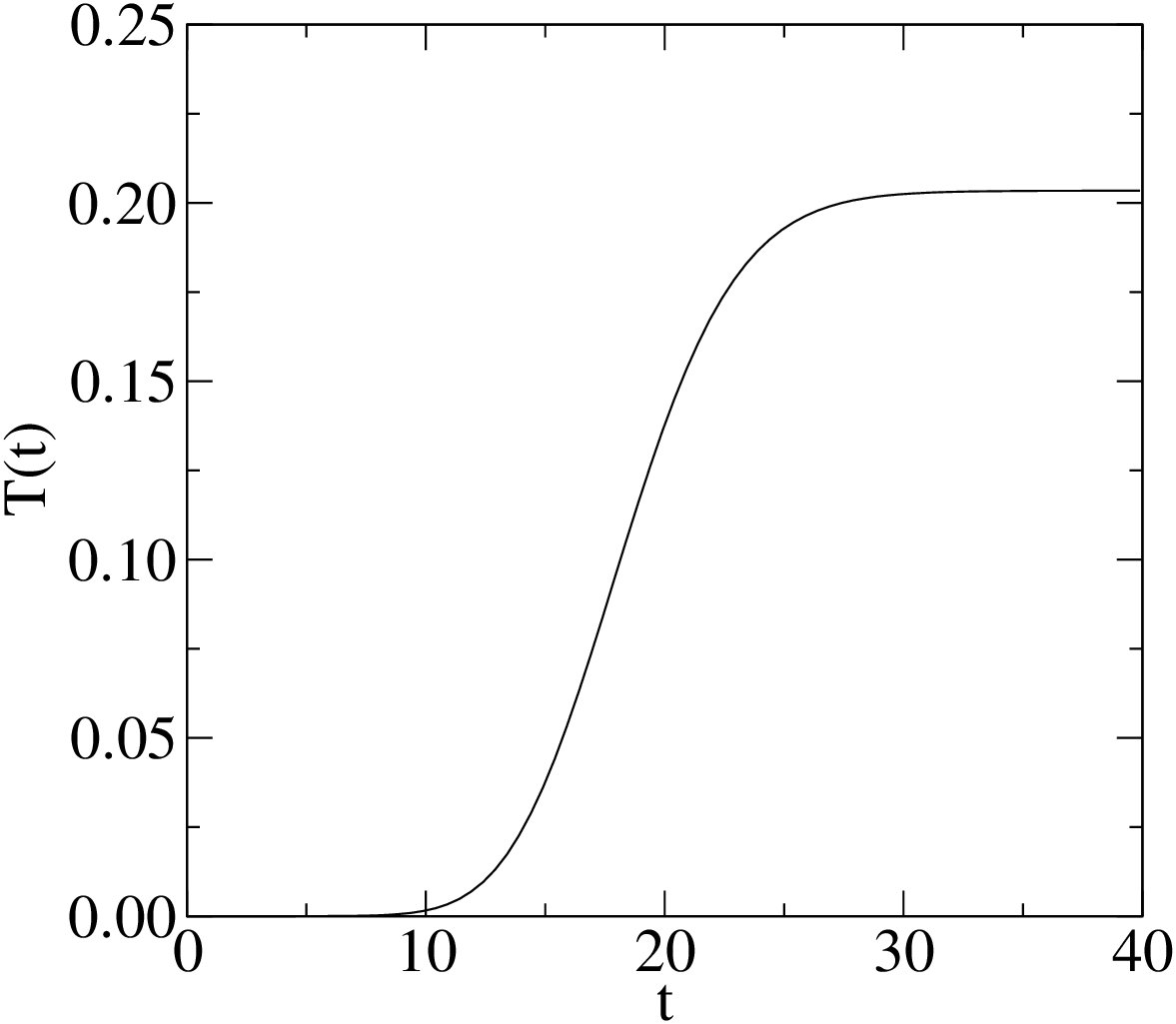}}
  \caption{\label{norm_1} Transmitted norm $T(t)$ for the case $K =
  1.06$ and for the reference potential.}
\end{figure}

We display in Figs.~\ref{norm_1} to \ref{norm_5}, the transmitted norm
of the wave $\Psi_l$. The transmitted norm is defined as a function of
time as
\begin{equation}
  T(t) = \int_5^{\infty} \left| \Psi_l( x, t ) \right|^2 \; dx.
  \label{tnorm}
\end{equation}
The lower bound in  the integral, $x = 5$, is chosen  to be far enough
away from  the barrier to  ensure no contamination of  the transmitted
wave by the barrier. We are also interested in the asymptotic value of
the transmitted  norm, $T(\infty)$. This observable  is also important
for comparisons  of different potentials.  Fig.~\ref{norm_1} shows the
case for the reference potential, $V(x) = e^{-x^2}$, with $K = 1.06$.
The asymptotic value  for this case is $T(\infty)  = 0.203$ confirming
the earlier  observation of 20\% transmission  for the wave  with $K =
1.06$.

\section{Spatial variations of the potential}
\label{space}

In this section, we investigate time-independent potentials which vary
with  respect   to  space.  As  we   are  discussing  time-independent
potentials, the time  variable is omitted from the  discussion for the
moment.

Given  the  same  starting   wave  packet  as  in  Section~\ref{base},
transmission through a barrier $V_1(x)$ will be less than that through
barrier $V_2(x)$  if $V_1(x)  > V_2(x)$, $\forall  x$. Details  of the
shape of  the incident  packet may change  this result. But  we assume
packets to be close to eigenstates, for which theorems bounding growth
and  curvatures  of  waves  in  relation to  the  potential  hold.  To
investigate  deviations from  this  estimate, we  compare results  for
potentials where $V_1(x) > V_2(x)$ for some $x$, and $V_1(x) < V_2(x)$
for other values.

This  is  achieved by  using  the  following  modulation to  our  base
potential,
\begin{equation}
  W(x) = \sigma e^{-2\omega x^2} \left[ \sin{\left( 11x
  \right) } \sin{
  \left( 13\sqrt{2} x \right) } \cos{ \left( 2\pi x \right) } \cos{
  \left( \frac{5x}{\sqrt{2}} \right) }
    +  \tau \sin{ \left( 3\sqrt{\pi} x \right) } \sin{
  \left( 7x \right) } \right],
  \label{potmod}
\end{equation}
where $\sigma$ is the strength  of the modulation and $\tau$, which is
weakly  dependent on $\sigma$,  is used  to cancel  the semi-classical
effect introduced by $W(x)$ (discussed below).

\subsection{Problem of a fair comparison}

\begin{figure}
  \scalebox{0.6}{\includegraphics*{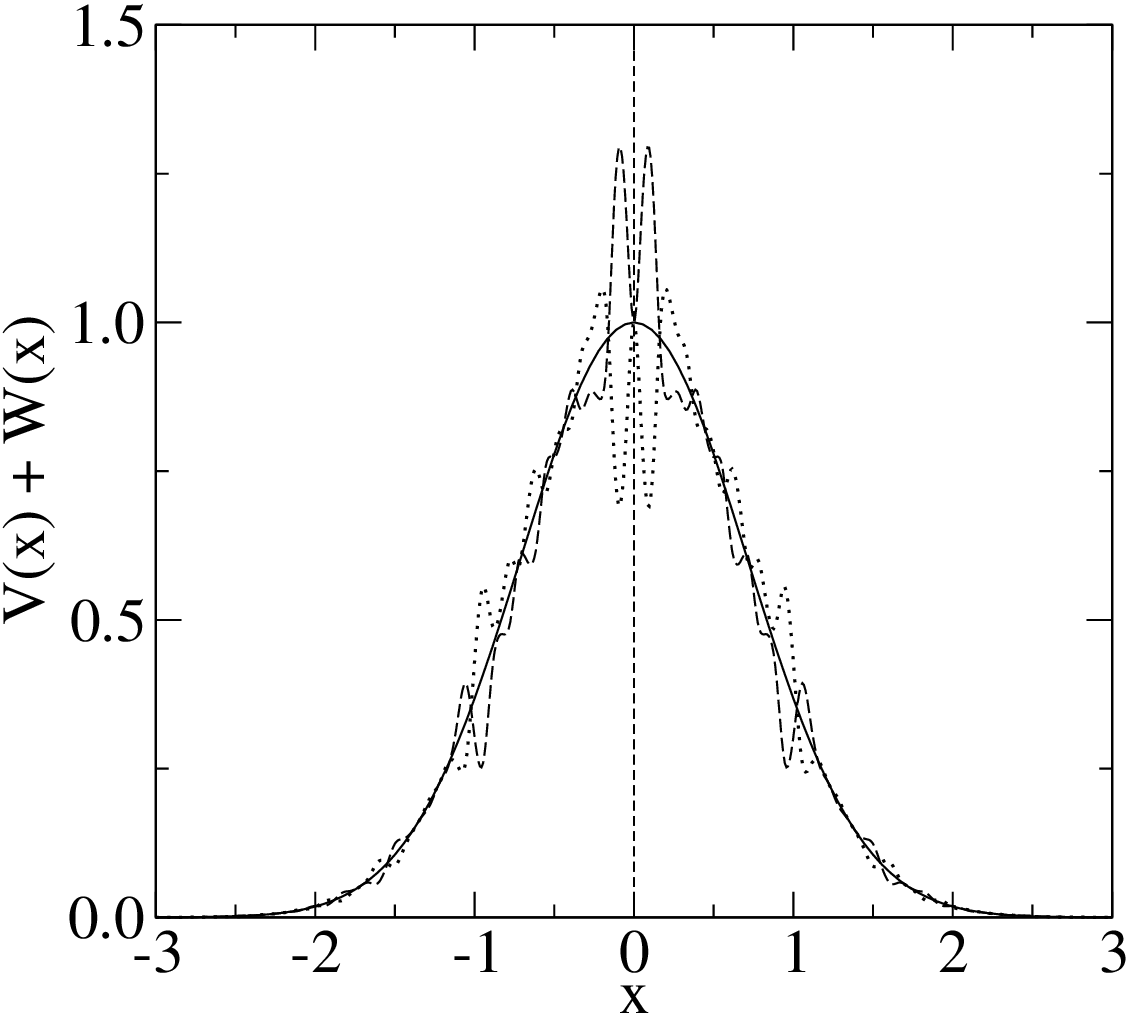}}
  \caption{\label{ears} The potentials used  for solving the TDSE. The
  base potential  ($W(x) = 0$) is  given by the solid  line, while the
  potentials for  $\sigma = 0.5$ (ears  up) and $\sigma  = -0.5$ (ears
  down) are given by the dashed and dotted lines respectively.}
\end{figure}

To achieve  a fair comparison  to our base  potential, we rely  on the
action integral
\begin{equation}
  \mathcal{A} = \int_{x_l}^{x_r} dx \sqrt{ 2 \left[ E - V(x) - W(x)
  \right] }\; ,
  \label{action}
\end{equation}
where $E$ is the energy of the packet and $x_l$ and $x_r$ are the left
and  right turning points  respectively. Of  course this  assumes that
there  are only  two such  turning points.  We compare  potentials for
which  $\mathcal{A}$ is  invariant. Fig.~\ref{ears}  shows  three such
potentials for $E = 0.571834$.
The  base  potential  $V(x)$  is  portrayed by  the  solid  line.  The
modulations  introduced by Eq.~(\ref{potmod})  correspond to  the case
$\sigma = 0.5$, $\tau = -0.111$ (dashed line, ``ears up'') and $\sigma
=  -0.5$, $\tau  = -0.132$  (dotted line,  ``ears down'').  The slight
difference  in  the  value  of  $\tau$ comes  from  the  condition  of
satisfying the semi-classical  action specified in Eq.~(\ref{action}).
If one requires that the average modulations vanish, i.e. $\mathcal{W}
= \int^{\infty}_{-\infty} W(x) dx =  0$, then $\tau = -0.118$; a value
not too  different from  the two  values given. In  fact, for  $\tau =
-0.111$, $\mathcal{W} = 8 \times  10^{-4}$, while for $\tau = -0.132$,
$\mathcal{W}  =  2  \times  10^{-3}$.  Thus the  modulations  we  have
introduced do not change  the semi-classical action and the associated
changes in the average of the potential are negligible.

\subsection{Results}

\begin{figure}
  \scalebox{0.6}{\includegraphics*{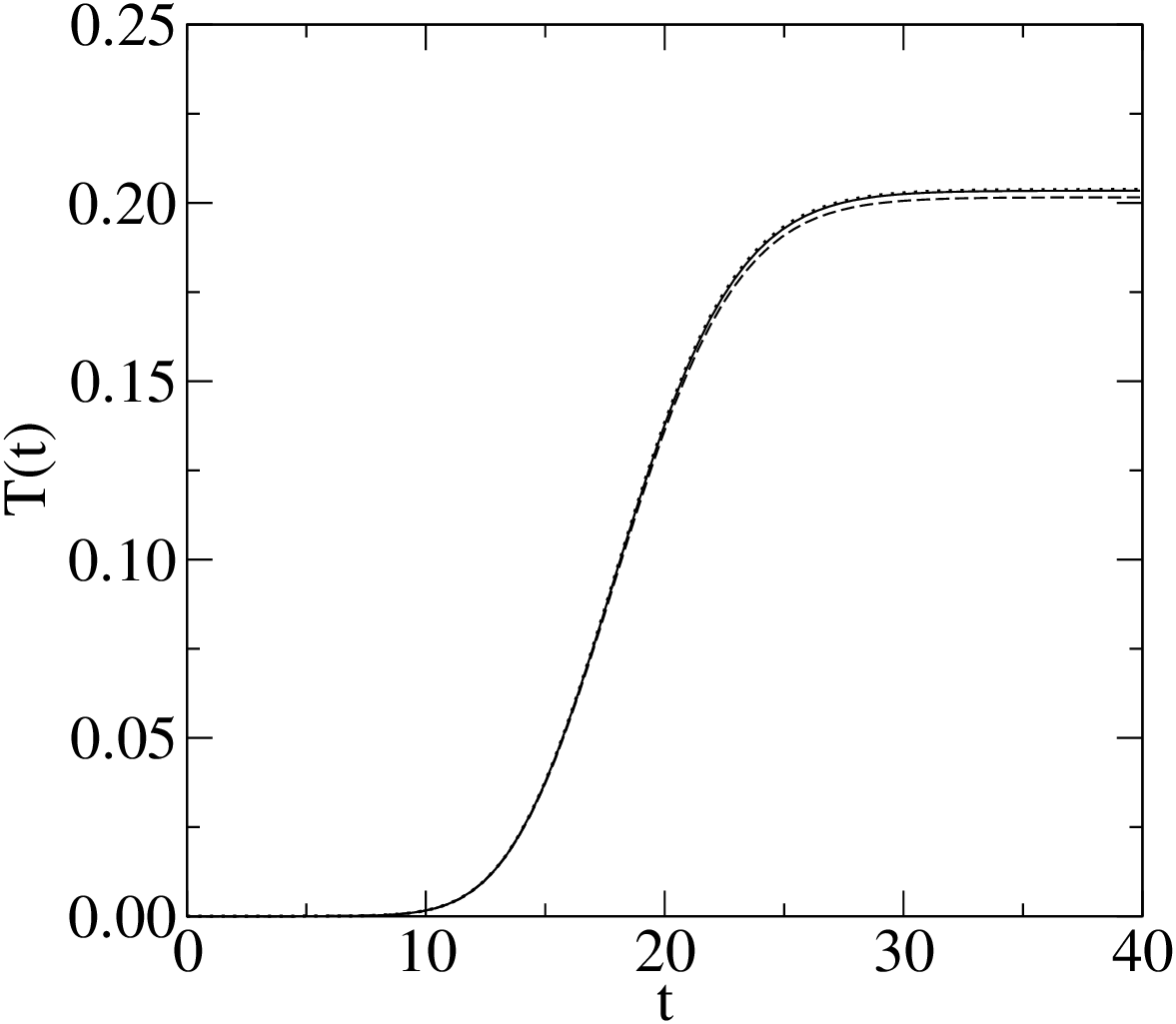}}
  \caption{\label{norm_2} Transmitted norms for  a wave packet with $K
  = 1.06$ incident  on the three potentials. The  curves correspond to
  the  results obtained for  the reference  (solid line),  ``ears up''
  (dashed line), and ``ears down'' (dotted line) potentials.}
\end{figure}

In Fig.~\ref{norm_2},  we display the transmitted norm  $T(t)$ for the
three potentials described  and for an incident wave  packet with $K =
1.06$.
Transmission  is  hardly affected  by  the  changes  to the  reference
potential. For the ``ears up'' potential $T(40) = 0.202$ while for the
``ears down'' it is 0.204. These  are to be compared with the value of
0.203 found using the reference potential. These changes, of the order
of a percent, are small  in comparison to the associated variations of
the potentials  from the  reference; the modulations  of which  are as
much as 30\%. That is especially so in the region of the ``ears''.

\begin{figure}
  \scalebox{0.6}{\includegraphics*{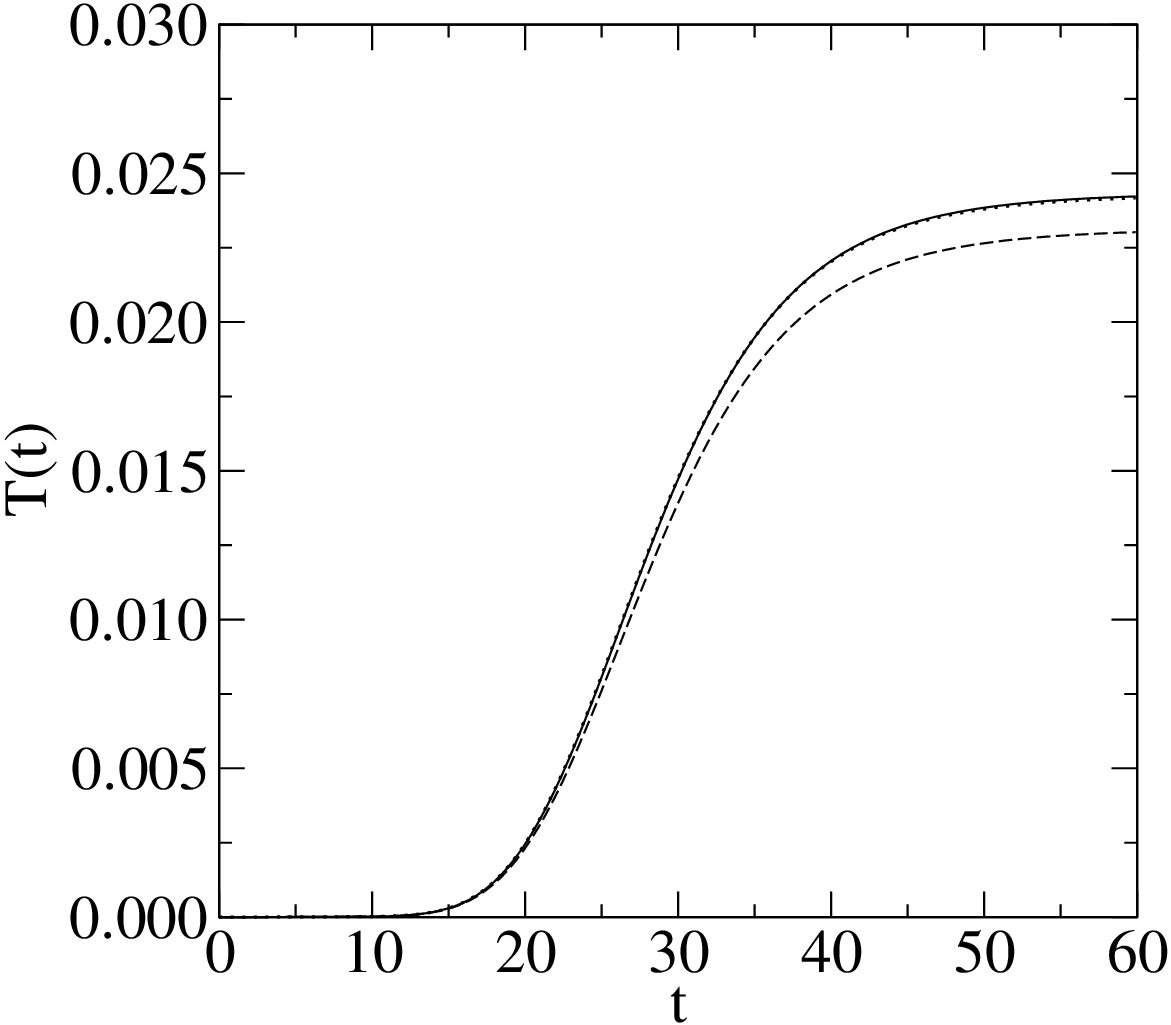}}
  \caption{\label{norm_3} As for Fig.~\ref{norm_2} but for $K = 0.6$.}
\end{figure}

Results have been obtained also for the same potentials but with lower
incident momenta. For  $K = 0.6$, hence $E  = 0.190034$, the condition
of fair comparison [Eq.~(\ref{action})] of potentials with modulations
defined by  Eq.~(\ref{potmod}) requires $\tau  = 0.050$ for  $\sigma =
0.5$ and  $\tau = -0.084$ for  $\sigma = -0.5$.  These parameters give
also the  ``ears up'' and  ``ears down'' potentials  respectively with
fluctuations on reference $\sim \pm 30\%$ in the ears. The results for
the transmitted norm in this case are displayed in Fig.~\ref{norm_3}.
For the  ``ears-up'' potentials at  this energy, the  transmission now
appears depleted as  the value of $T(\infty)$ in  this case is 0.0226.
But  there  is little  change  for  the  ``ears-down'' case  from  the
asymptotic  transmitted  norm obtained  for  the reference  potential,
0.0239.

\begin{figure}
  \scalebox{0.6}{\includegraphics*{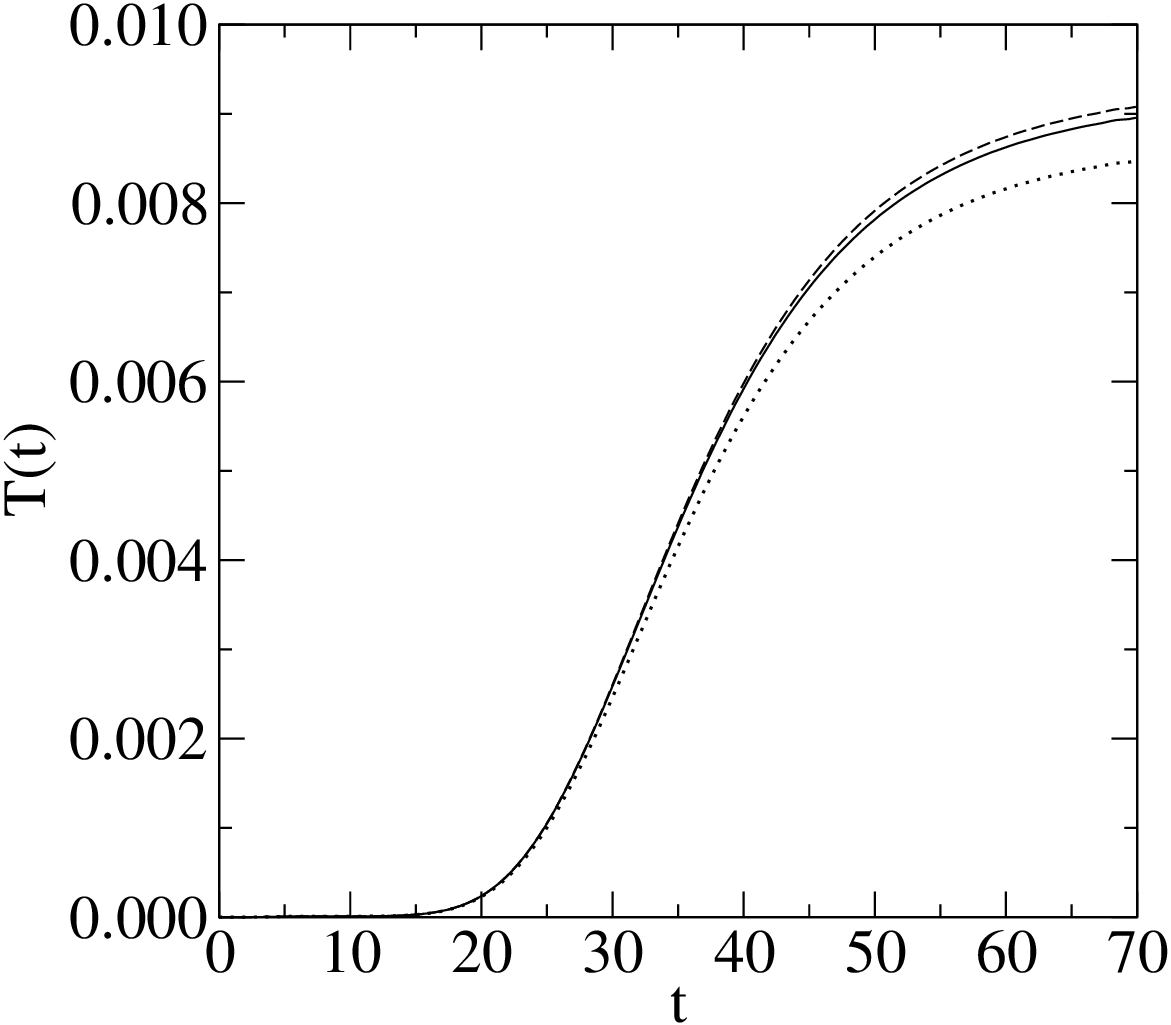}}
  \caption{\label{norm_4} As for Fig.~\ref{norm_2} but for $K = 0.42$.}
\end{figure}

A different picture occurs for the case $K = 0.42$, or $E = 0.098234$.
This is  displayed in Fig.~\ref{norm_4}  for the same  potentials used
previously.
For the lowest  momentum considered, there is a  slight enhancement in
the  transmission from  the  ``ears-up'' potential  and a  significant
depletion  with  the  ``ears-down''  potential. This  is  the  reverse
situation to that with $K = 0.6$.

\begin{figure}
  \scalebox{0.6}{\includegraphics*{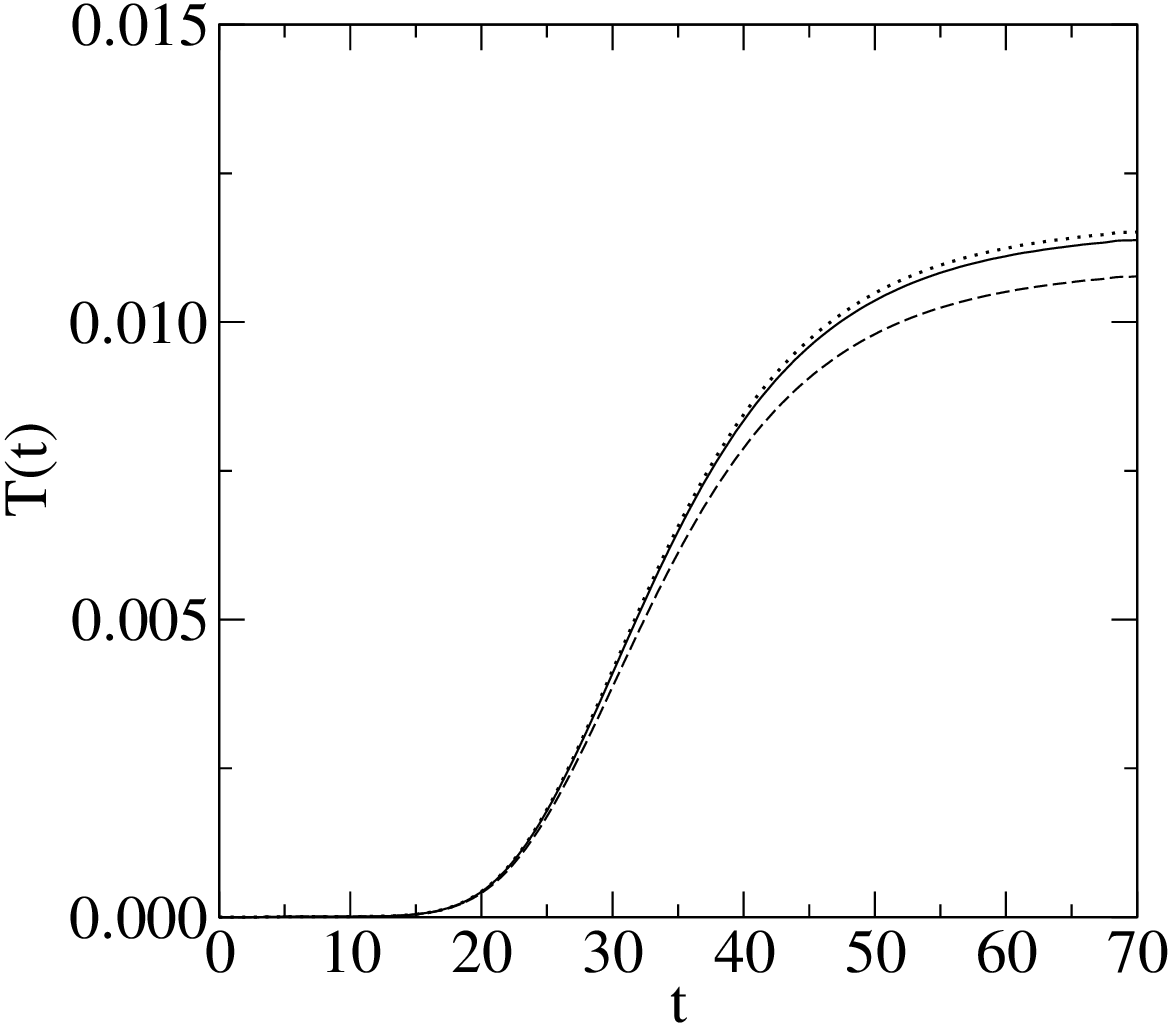}}
  \caption{\label{norm_5} As for Fig.~\ref{norm_2} but for $K = 0.46$
  and with the modulating potential $W(x)$ given by Eq.~(\ref{mod2}).}
\end{figure}

The  modulating potential  $W(x)$ in  [Eq.~(\ref{potmod})] is  not the
only form that we have considered. Another is
\begin{equation}
  W(x)  =  \sigma e^{-2\omega  x^2}  \left[  \sin{\left( 17x  \right)}
  \sin{\left(  13\sqrt{2}x  \right)} \cos{\left(  \frac{11x}{\sqrt{2}}
  \right)} \cos{\left( 7\pi x \right)} + \tau \sin{\left( 3\sqrt{\pi}x
  \right)} \sin{\left( 7x \right)} \right].
  \label{mod2}
\end{equation}
The  results  for  the  transmitted  norm  with  this  modulation  are
displayed in Fig.~\ref{norm_5} for $K  = 0.46$, $\sigma = \pm 0.5$ and
$\tau =  0.141$ and  $0.028$, respectively for  the sign  of $\sigma$.
While there
appears  to be little  change in  transmission with  the ``ears-down''
potential there is some depletion with the ``ears-up'' case.

It  is  noteworthy  that,  with  any of  these  modifications  to  the
reference  potential,  the  effect  on the  transmission  is  minimal;
changes being at most of the  order of 10\%. This is not sufficient to
explain  the observed  depletion  of fusion  rates  below the  Coulomb
barrier.

\section{Fluctuations in time}
\label{time}
As the spatial fluctuations are unlikely to be the source of the large
loss of fusion that is  observed experimentally, we turn our attention
to time-dependent fluctuations on  the base potential. We assume those
fluctuations are of the form
\begin{equation}
  v(t) = 1 + \gamma \cos{\left( \Omega_c t \right)} \sin{ \left(
  \Omega_s t \right) }.
  \label{tfluct}
\end{equation}
We take $\gamma  = 0.2$ while $\Omega_c$ and  $\Omega_s$ are chosen at
random  with  uniform  distributions  varying  between  0  and  5  for
$\Omega_c$ and between $-5$ and 5 for $\Omega_s$. These parameters are
then sampled allowing  for a good simulation of  the chaotic character
of $(v(t) - 1)$.

We  start again with  our initial  wave packet,  Eq.~(\ref{pack}). The
fine  structures  in  the  packet  experience  the  weakly  correlated
components  of the  oscillating fluctuations  and  so we  need not  be
concerned about  any time periodicity in $V(x,t)$.  The sampling space
is one  of 200 to  500 potentials (independent choices  for $\Omega_c$
and $\Omega_s$)  and we  use the asymptotic  value of  the transmitted
norm  $T(\infty)$  as the  measure  of  effect  of the  time-dependent
potentials  in comparison  to that  from the  reference  potential for
which $v(t) = 1$. The latter we designate as $T_R(\infty)$.

\begin{figure}
  \scalebox{0.6}{\includegraphics*{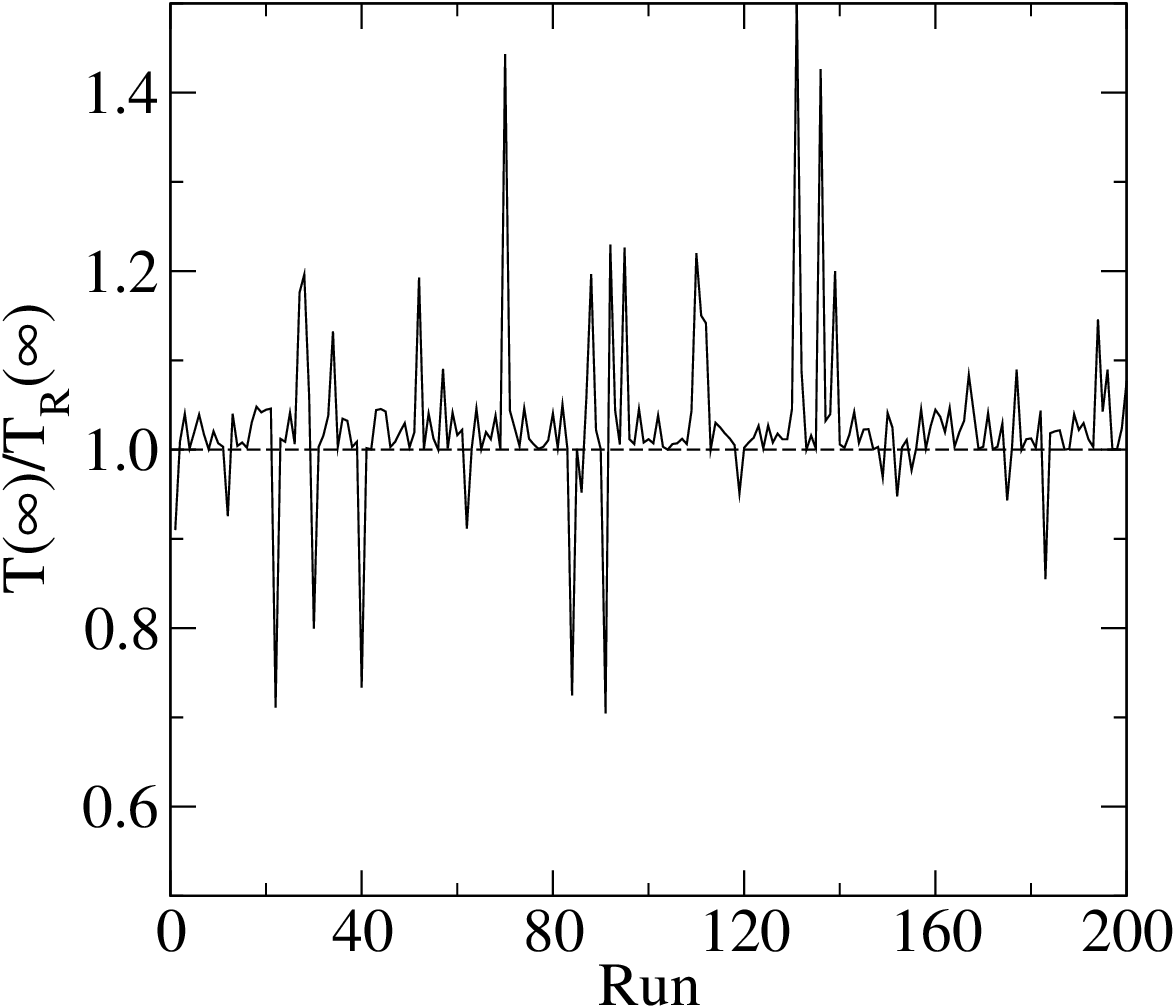}}
  \caption{\label{runs}  Ratio of  $T(\infty)$  against the  reference
  value ($T_R(\infty)  = 0.0239$) for  the runs made using  the random
  pair   of  $\left\{  \Omega_c,   \Omega_s  \right\}$   specified  in
  Eq.~(\ref{tfluct}). For  these calculations, $K=0.6$. The  line is a
  guide to the eye.}
\end{figure}

\begin{figure}
  \scalebox{0.6}{\includegraphics*{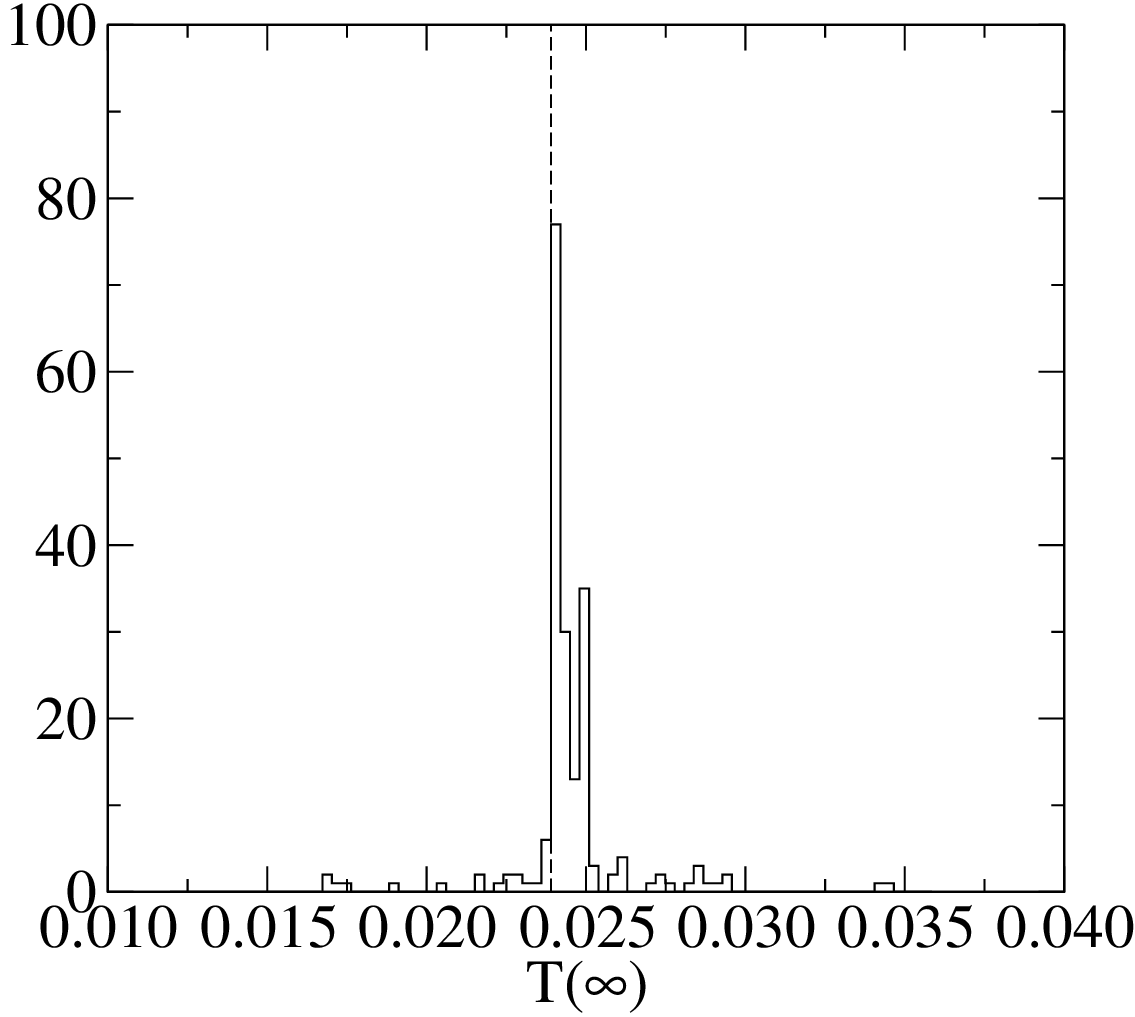}}
  \caption{\label{hist1} Distribution of values of $T(\infty)$ for the
  runs  shown in  Fig.~\ref{runs},  for which  $K=0.6$. The  reference
  value $T_R(\infty) = 0.0239$ is indicated by the dashed line.}
\end{figure}

\begin{figure}
  \scalebox{0.6}{\includegraphics*{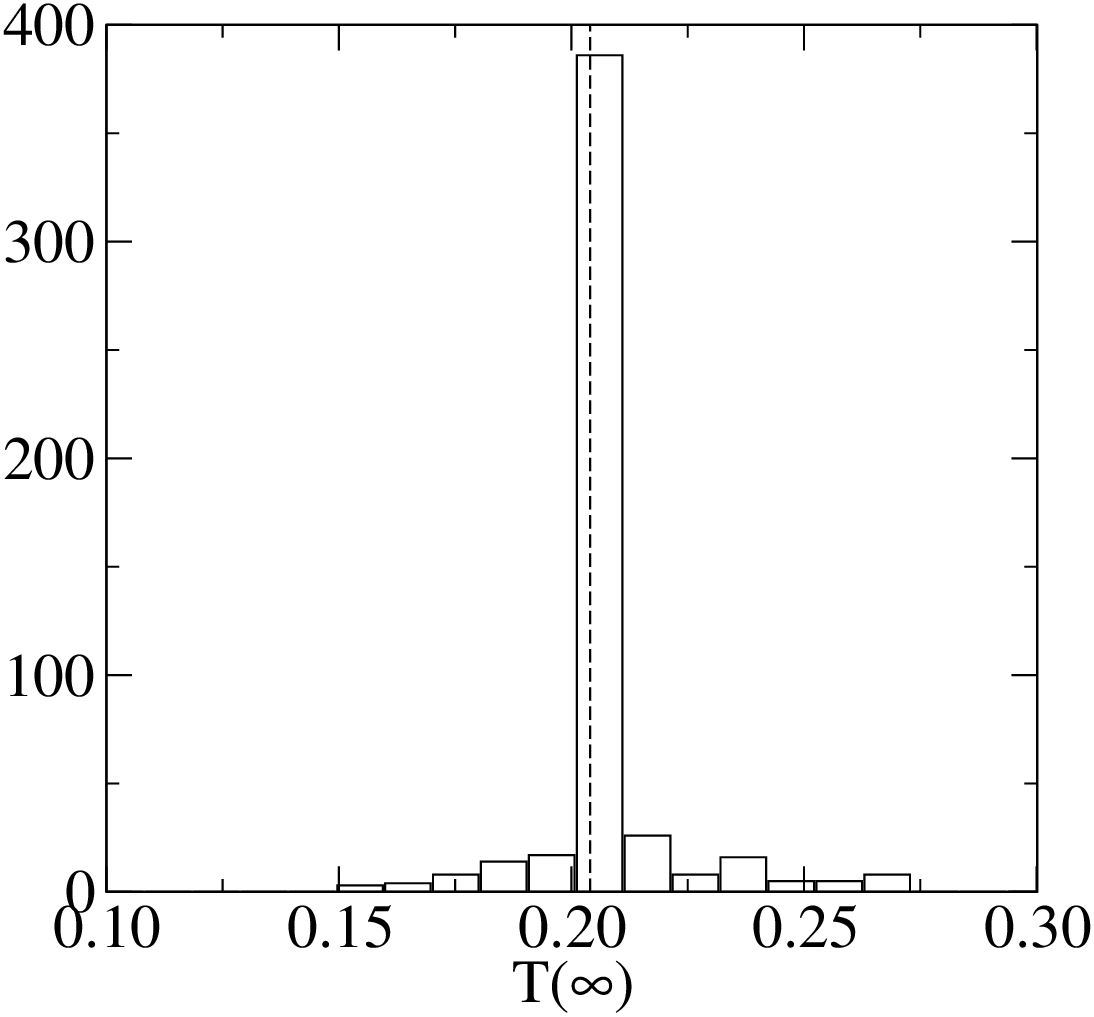}}
  \caption{\label{hist3} As  for Fig.~\ref{hist1}  but for $K  = 1.06$
    and using a  sample of 500 potentials. The  value of $T_R(\infty)$
    for the reference potential is 0.2039.}
\end{figure}

Displayed in Fig.~\ref{runs}  is the ratio of $T(\infty)/T_R(\infty)$,
obtained from 200 runs for a packet with initial momentum $K = 0.6$.
The reference value for this  case is $T_R(\infty) = 0.0239$. For most
of  the pairs  sampled, the  transmission  is close  to the  reference
potential.  But there  are  a  few instances  where  the tunneling  is
greatly  enhanced as  well  as  others where  it  is greatly  reduced.
Variations of  as much  as 50\% occur.  The distribution of  values of
$T(\infty)$ for $K = 0.6$ is displayed in Fig.~\ref{hist1}.
As indicated in Fig.~\ref{runs},  the introduction of time fluctuating
potentials  increases slightly  the  value of  $T(\infty)$ on  average
indicating that the  transmission is enhanced, if only  a little. That
is a general feature we find for many conditions and one such is shown
in Fig.~\ref{hist3}. Therein the histogram for runs with
$K = 1.06$ (sample size of  500) are displayed. The slight increase of
$T(\infty)$ on average is evident again.

\section{3-dimensional model}
\label{radial}

\subsection{Formalism for a more realistic model}

\begin{figure}
  \scalebox{0.6}{\includegraphics*{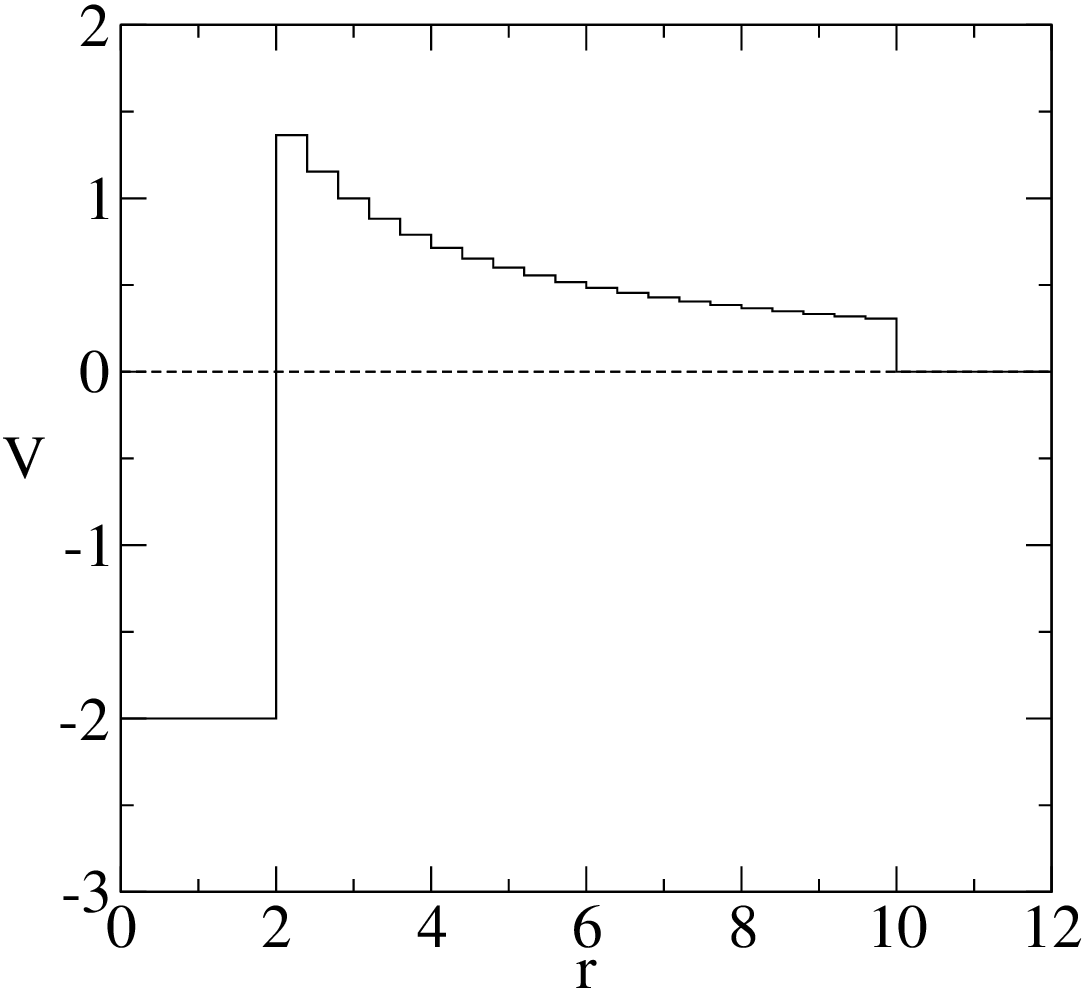}}
  \caption{\label{radpot} The  radial potential $V_{\text{sqC}}(r)$ in
  the 3D problem.}
\end{figure}

We  now consider  3-dimensional eigenstates  of the  Hamiltonian  $H =
-\partial^2/(2\partial     r^2)     +    V_{\text{sqC}}(r)$,     where
$V_{\text{sqC}}(r)$  is   an  approximation  to   the  central  radial
potential  for a  realistic nucleus.  The potential  is built  using a
series of square wells  to approximate an attractive nuclear potential
plus a Coulomb  barrier. In our dimensionless units,  the nuclear part
is taken as a square well with depth $U = -2$ with range $R = 2$ while
the Coulomb interaction  is approximated by a staircase  with 20 steps
between $r = 2$  and $r = 10$, with no repulsion  beyond $r = 10$. The
center of the $n^{\text{th}}$ step is $r_n  = 2.2 + 0.4(n - 1)$ and so
to  construct a  Coulomb barrier  with strength  3 in  our  units, the
height of the $n^{\text{th}}$ step  is $3/r_n$. The potential is shown
in Fig.~\ref{radpot}.

We shall consider  tunneling using this potential for  energies in the
range $0.1 < E < 0.7$,  equating roughly from a third of the outermost
step to  the height of  the middle (sixth)  step, i.e. about  half the
maximum  barrier.  This  choice  is   dictated  by  a  need  to  study
sub-Coulomb  processes  as well  as  a  need  to minimize  effects  of
discontinuities arising from the  highest steps in the staircase. Also
we consider  $s$-wave scattering only as  this simplifies calculations
using square wells. For a given  incident momentum $K$ and energy $E =
K^2/2$, the momentum inside the potential is $k = \sqrt{K^2 - 2U}$. We
normalize the solution by a  unit derivative for the innermost part of
the wave, namely $\sin{(kr)}/k$, so that we can define the square norm
for ``presence in  the attractive well'' as $\mathcal{N}  = R/(2k^2) -
\sin{(2kR)}/(4k^3)$. With the incident wave function given by
\begin{equation}
\Psi = A \left[ e^{-iKr} + B e^{iKr} \right],
\end{equation}
matching the wave function and its logarithmic derivative at each step
boundary gives  the parameters $A$ and  $B$. At each  energy, the test
for  unitarity, $\left| B  \right| =  1$, is  made to  check numerical
accuracy.

The relevant measure  we define as the transmission  index; a quantity
given by  $\mathcal{T} = \mathcal{N}/\left|  A \right|$. This  one can
view  as an ``inside  square norm  per unit  (incoming) flux''.  For a
given energy  $E$, it is  also possible to  find both outer  and inner
turning    points,   and    the    corresponding   classical    action
[Eq.~(\ref{action})].

\subsection{Results}

\begin{figure}
  \scalebox{0.6}{\includegraphics*{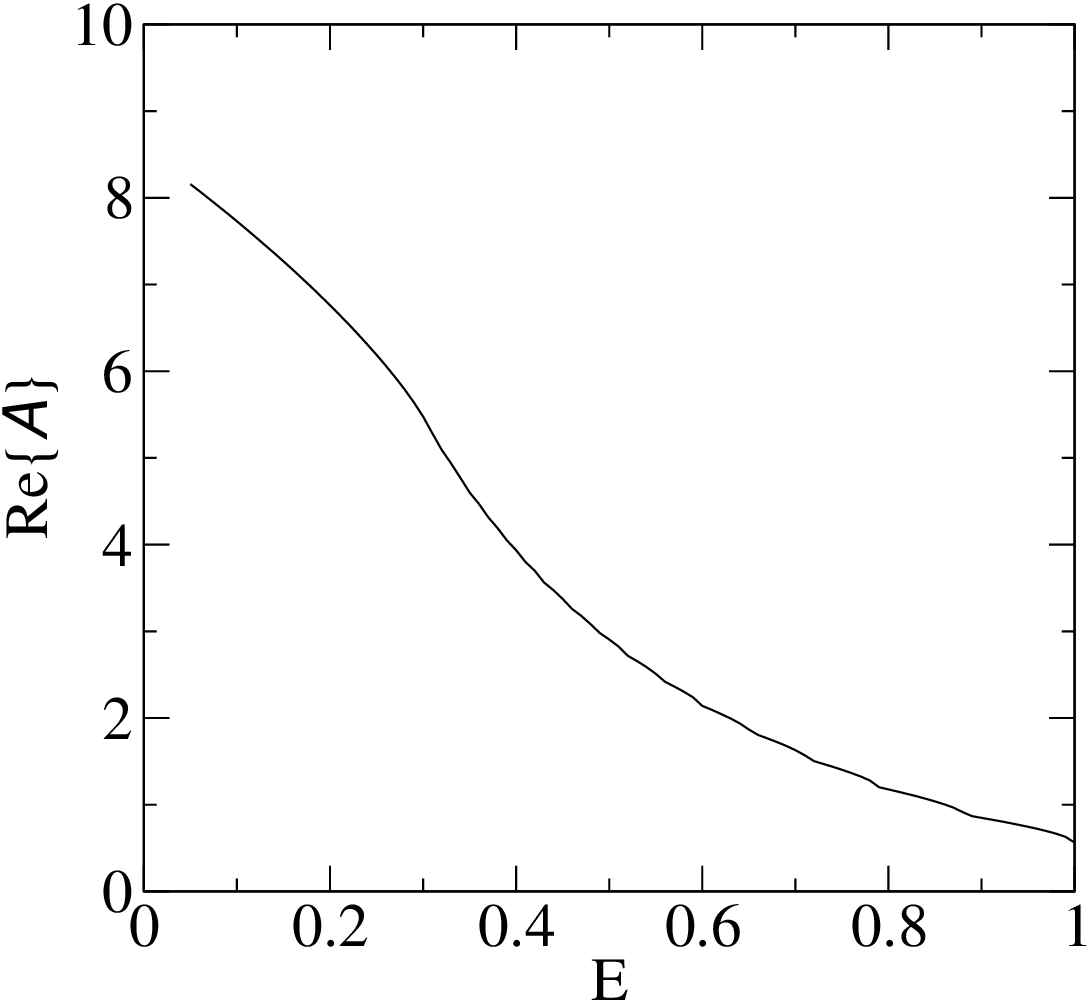}}
  \caption{\label{actfig} Energy variation  of the reference classical
    action for tunneling under the potential $V_{\text{sqC}(r)}$.}
\end{figure}

\begin{figure}
  \scalebox{0.6}{\includegraphics*{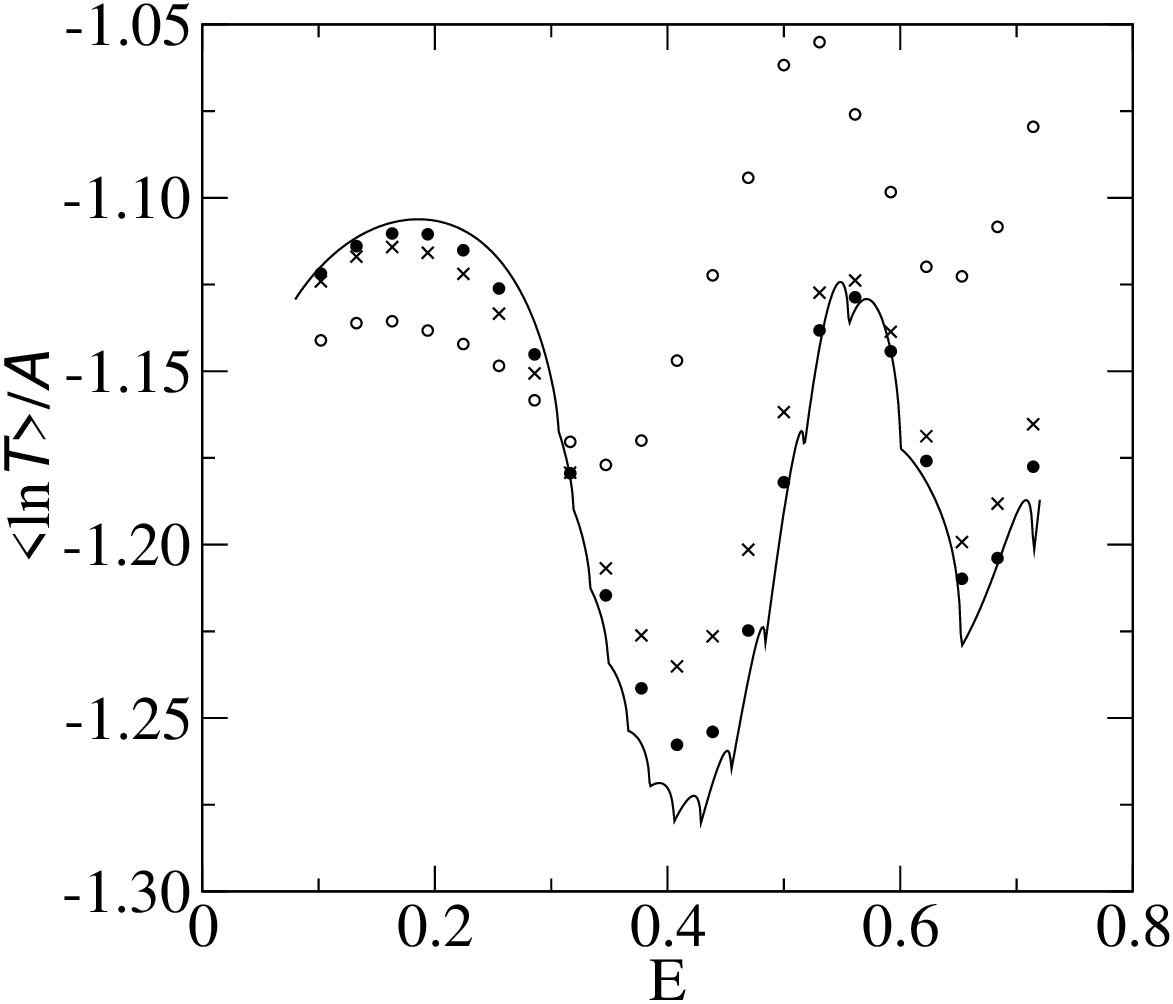}}
  \caption{\label{radrat}    Ratio    $\left\langle   \ln{\mathcal{T}}
  \right\rangle/\mathcal{A}$ as a function  of energy. The solid curve
  was obtained by using  the reference potential $V_{\text{sqC}}$. The
  closed circles depict the result found by using a fluctuation energy
  of 0.0929  while the open circles  correspond to use of  an energy 3
  times  greater.  The  crosses  are  the results  found  on  using  a
  fluctuation set at 10\% of the highest strength value.}
\end{figure}

Solution  of the tunneling  problem with  the 3-D  staircase potential
give the classical action displayed in Fig.~\ref{actfig} as a function
of $E$.
While  the  discontinuities  in   the  staircase  potential  make  the
derivative  of $\mathcal{A}$  discontinuous,  $\mathcal{A}$ itself  is
continuous  and the  derivative  does not  jump  significantly. It  is
expected  that  $\ln{\mathcal{T}}$  will  be roughly  proportional  to
$\mathcal{A}$. This is confirmed in Fig.~\ref{radrat}, which shows the
ratio of $\ln{\mathcal{T}}$ to $\mathcal{A}$ as a function of $E$.
Despite  oscillations and  a fine  structure,  the latter  due to  the
discontinuities in  the staircase potential  approximating the Coulomb
interaction, an average value for the  ratio of $-1.2$ to within a few
percent  is  observed for  all  of  the  calculations made,  including
fluctuations that are described next.

To incorporate  fluctuations we introduce a  random modification $w_n$
of the $n^{\text{th}}$ step which  raises or lowers the height of that
step. The sampling is from a uniform distribution with typically
\begin{equation}
\frac{3}{r_5} - \frac{3}{r_4} < w_n < \frac{3}{r_4} - \frac{3}{r_5}
\text{ where } \frac{3}{r_4} - \frac{3}{r_5} = 0.0929\, .
\end{equation}
We also  allow for  an increase or  a decrease  of this range  by some
factor. But  introducing such  fluctuations no longer  guarantees that
the staircase  potential will  be monotonic and  it may allow  for the
propagation  of  the wave  at  positive  local  energies above  a  few
intermediate  steps.  This will  give  more  than  two turning  points
although the number will always be even.

We  must also  consider the  fairness criterion  of Eq.~(\ref{action})
when invoking  fluctuations. We do  this for each  perturbed staircase
potential  $V_p$ by calculating  the classical  action $\mathcal{A}'$,
taking into account the possibility  of the existence of more than two
turning points because of the  (possible) presence of local minima. We
compare  $\mathcal{A}'$  to  that  ($\mathcal{A}$)  of  the  reference
staircase  potential  and reject  any  $V_p$  from  our sample  if  $|
\mathcal{A}'/\mathcal{A} - 1 | >  10^{-4}$. As a check, we verify that
our results do not depend  on any variation in the tolerance. Finally,
we average $\ln{\mathcal{T}}$ over  the subset of perturbed potentials
that meet the selection criterion.

For each  energy $E$,  we sample between  $10^5$ and  $10^6$ staircase
potentials in  the fluctuation space. However,  the fairness criterion
is  severe: only  a few  thousand, or  even a  few  hundred, staircase
potentials are  retained. The acceptance rate,  an increasing function
of the  tolerance as  well as  a decreasing function  of the  range of
fluctuations, also is a  decreasing function of the energy. Typically,
the number of accepted potentials is  five times smaller for $E = 0.7$
than it is for $E = 0.1$. Possibly that is due to the local momenta $k
= \sqrt{ 2\left[ E - V_p(r)  \right] } \to 0$ with a larger derivative
as $E$  increases, which in turn  would lead to a  larger variation in
the classical  action. Note  also, that for  such numbers  of accepted
potentials, we  also verify that the  average of the  logarithm of the
transmission   index  $\left\langle   \ln{\mathcal{T}}  \right\rangle$
barely  differs  from  the   logarithm  of  the  average  transmission
$\ln{\left\langle  \mathcal{T} \right\rangle }$.  A typical  result is
shown  in   Fig.~\ref{radrat}  where   the  dots  correspond   to  the
fluctuation range  0.0929 and  the open circles  to one that  is three
times larger. An interesting  phenomenon is that fluctuations decrease
the transmission  at lower energies  while they increase it  at higher
energies. The transition between these two regimes occurs near $E \sim
0.3$.  But this  would not  seem to  be a  significant reason  for the
depletion or enhancement of tunneling,  as long as the fluctuations do
not  exceed 10\%  of the  potential.  That result  corresponds to  the
crosses in Fig.~\ref{radrat}. We verified this result numerically also
for other  parameter sets within the  model. It may  be concluded that
within our present understanding  of friction, the lower than expected
fusion cross  sections cannot be explained by  the processes discussed
either with the 1D or the 3D models.

\section{Discussion and conclusions}
\label{tisdone}

We  have considered  various cases  of  tunneling in  a fully  quantal
approach, changing  the base potential  in our model by  adding either
space-dependent or time-dependent fluctuations  to see if there is any
enhancement in the transmission of  the packet beyond the barrier. For
the  cases of  the space-dependent  fluctuations, the  induced changes
were made such that the classical action was invariant so allowing for
a  fair comparison  of the  results obtained  with those  of  the base
(reference)  potential.  The  effects  of  those changes  seem  to  be
momentum dependent. For a  high incident momentum, corresponding to $K
= 1.06$, there  is very little change in the  transmission of the wave
through the barrier. For much lower momenta, particularly the case for
$K = 0.42$, there is  change, with the ``ears up'' potential producing
a reduction in the transmission.  The same occurs at this momentum for
an effectively random change in the modulating potential. But the size
of the changes in the transmission are not very large, typically $\sim
10\%$ and given that we  can produce both enhancement and depletion by
such (relatively large)  changes in the barrier, we  conclude that the
cause   of  the   large  ($\sim   50\%$)  loss   of   fusion  observed
experimentally is  unlikely to be  solely, or even largely,  caused by
changes in the transmission due to space fluctuations in the barrier.

By perturbing  the barrier  with time-dependent oscillations,  we have
been able to produce a  small systematic increase in the transmission.
Yet with our  sample over a large number of  potentials and at various
incident  momenta we were  only able  to find  small numbers  of cases
where  the  transmission  was   either  greatly  enhanced  or  greatly
diminished. The enhancement may correspond to a situation where fusion
is also enhanced and vice-versa.  However, the number of such cases is
relatively few, and  the average of all lead  to small enhancements in
transmission for all momenta.

We  also considered a  more realistic  three-dimensional case  using a
potential  made up  of  a  series of  square  wells approximating  the
nuclear plus  Coulomb potentials  of the nucleus.  This allowed  us to
treat  the  problem  analytically   and  so  consider  the  effect  of
transmission more closely. Again we allowed some fluctuations into the
system,  this  time  by   small  perturbations  to  the  square  wells
constructing the  Coulomb barrier. As with  the one-dimensional cases,
we observed only small changes to the results of transmission obtained
for the reference radial  potential. There were no significant changes
that might produce the observed lack of fusion.

In  all cases,  there is  one  overriding consideration:  there is  no
evidence for a ``friction'' related to a Langevin process with complex
time.  By considering  the  full  quantal TDSE  no  such effects  with
classical analogues,  or those involving  complex time, are  needed to
produce changes  to the observed  transmission. But those  changes are
not significant enough to indicate that the source of the depletion of
fusion  rates in  heavy-ion reactions  at extreme  sub-Coulomb barrier
energies comes from changes in tunneling.

\begin{acknowledgments}
  We thank J. M. Luck for stimulating discussions during the course of
  this  work.  We thank  also  J.-P. Delaroche,  D.  J.  Hinde and  M.
  Dasgupta for helpful comments. Also, one of us (B.G.G.) acknowledges
  the hospitality  of the Australian  National University and  for the
  motivation for this project.
\end{acknowledgments}

\bibliography{tunnel_new}

\end{document}